\RequirePackage{ifpdf}
\newcommand{\commondocopts}{letterpaper,aps,pra,10pt,superscriptaddress,showpacs,floats,twocolumn,nofootinbib} 
\ifpdf
  \documentclass[\commondocopts,pdftex]{revtex4-1}
  \usepackage{cmap}
\else
  \documentclass[\commondocopts,ps2pdf]{revtex4-1}
\fi
\usepackage[latin1]{inputenc}
\usepackage[T1]{fontenc}
\usepackage{amsmath,amssymb}

\usepackage{graphicx,color}
\usepackage{hyperref}

\definecolor{rltred}{rgb}{0.75,0,0}
\definecolor{rltgreen}{rgb}{0,0.5,0}
\definecolor{rltblue}{rgb}{0,0,0.75}
\hypersetup{colorlinks,%
hypertexnames=true,%
pdfauthor={Renate Pazourek, Johannes Feist, Stefan Nagele, Emil Persson, Barry I. Schneider, Lee A. Collins, Joachim Burgd\"orfer},%
pdftitle={Universal features in sequential and nonsequential two-photon double ionization of helium},%
pdfview=FitH,%
pdfstartview=FitH,%
pdfpagemode=UseNone,%
bookmarksopen=true,%
bookmarksnumbered=true,%
pdfhighlight=/I,%
linkcolor=rltred,%
citecolor=rltgreen,%
linktocpage=true}

\begin{document}

\renewcommand{\equationautorefname}{Eq.}
\renewcommand{\figureautorefname}{Fig.}
\newcommand{\refeq}[1]{\hyperref[#1]{\equationautorefname~(\ref*{#1})}}

\newcommand{\Int}{\int\limits}
\newcommand{\sumint}[1]{\sum\mkern-25mu\Int_{#1}\;}
\newcommand{\E}{\mathcal{E}}

\newcommand{\etal}{\emph{et~al.}\ }
\newcommand{\ie}{i.e., }
\newcommand{\Schro}{Schr\"o\-din\-ger }
\newcommand{\eg}{e.g., }
\newcommand{\cf}{cf.~}

\newcommand{\expval}[1]{\langle#1\rangle}
\newcommand{\abs}[1]{\left|#1\right|}

\newcommand{\bra}[1]{\langle#1|}
\newcommand{\ket}[1]{|#1\rangle}
\newcommand{\braOket}[3]{\langle#1|#2|#3\rangle}

\newcommand{\unitv}[1]{\mathbf{\hat{#1}}}

\newcommand{\cvec}[1]{\mathbf{#1}}
\newcommand{\op}[1]{\mathrm{\hat{#1}}}
\newcommand{\vecop}[1]{\cvec{\hat{#1}}}
\newcommand{\eqcomma}{\,,}
\newcommand{\eqstop}{\,.}
\newcommand{\ed}{\,}

\newcommand{\dt}{\dd t}
\newcommand{\dE}{\dd E}
\newcommand{\dd}{\mathrm{d}}
\newcommand{\hw}{\hbar\omega}

\newcommand{\submax}{\mathrm{max}}
\newcommand{\Lmax}{L_\submax}
\newcommand{\lonemax}{l_{1,\submax}}
\newcommand{\ltwomax}{l_{2,\submax}}

\newcommand{\ev}{\,\mathrm{eV}}
\newcommand{\eV}{\ev}
\newcommand{\au}{\,\mathrm{a.u.}}
\newcommand{\Wcm}{\,\mathrm{W}/\mathrm{cm}^2}
\newcommand{\as}{\,\mathrm{as}}
\newcommand{\fs}{\,\mathrm{fs}}
\newcommand{\He}{\mathrm{He}}
\newcommand{\Hep}{\He^+}

\newcommand{\Tp}{T}
\newcommand{\Etot}{E_\mathrm{total}}
\newcommand{\DE}{\Delta E}
\newcommand{\PDI}{P^{DI}}
\newcommand{\pz}[1]{\op{p}_{z,#1}}
\newcommand{\G}{\mathcal{G}}
\newcommand{\Gsinsq}{\G_{\sin^2}}
\newcommand{\Eni}{E_{ni}}
\newcommand{\Efn}{E_{fn}}

\newcommand{\expcth}{\expval{\cos\theta_{12}}}

\title{Universal features in sequential and nonsequential two-photon double ionization of helium}

\author{R.~Pazourek}
\email{renate.pazourek@tuwien.ac.at} 
\affiliation{Institute for Theoretical Physics, 
             Vienna University of Technology, 1040 Vienna, Austria, EU}

\author{J.~Feist} 
\affiliation{ITAMP, Harvard-Smithsonian Center for Astrophysics, 
		     Cambridge, Massachusetts 02138, USA}
\affiliation{Institute for Theoretical Physics, 
             Vienna University of Technology, 1040 Vienna, Austria, EU}

\author{S.~Nagele}
\affiliation{Institute for Theoretical Physics, 
             Vienna University of Technology, 1040 Vienna, Austria, EU}
             
\author{E.~Persson}
\affiliation{Institute for Theoretical Physics, 
             Vienna University of Technology, 1040 Vienna, Austria, EU}

\author{B.~I.~Schneider}
\affiliation{Office of Cyberinfrastructure/Physics Division, National Science Foundation,
			 Arlington, Virginia 22230, USA}

\author{L.~A.~Collins}
\affiliation{Theoretical Division, 
             Los Alamos National Laboratory, Los Alamos, New Mexico 87545, USA}

\author{J.~Burgd\"orfer}
\affiliation{Institute for Theoretical Physics, 
             Vienna University of Technology, 1040 Vienna, Austria, EU}

\date{\today}

\begin{abstract}
We analyze two-photon double ionization of helium in both the nonsequential ($\hw<I_2\approx54.4\ev$) and sequential ($\hw>I_2$) regime.
We show that the energy spacing $\DE=E_1-E_2$ between the two emitted electrons provides the key parameter that controls both the energy and the angular distribution and reveals the universal features present in both the nonsequential and sequential regime.
This universality, \ie independence of $\hw$, is a manifestation of the continuity across the threshold for sequential double ionization.
For all photon energies, the energy distribution can be described by a universal \emph{shape function} that contains only the spectral and temporal information entering second-order time-dependent perturbation theory.
Angular correlations and distributions are found to be more sensitive to the value of $\hw$.
In particular, shake-up interferences have a large effect on the angular distribution. 
Energy spectra, angular distributions parameterized by the anisotropy parameters $\beta_j$, and total cross sections presented in this paper are obtained by fully correlated time-dependent \emph{ab initio} calculations.

\end{abstract}
\pacs{32.80.Rm, 32.80.Fb, 42.50.Hz}

\maketitle

\section{Introduction}\label{sec:intro}
Two-photon double-ionization (TPDI) of helium is a prototype process for the study of electron correlation. The advent of ultrashort and intense light sources with sufficient photon flux \cite{HenKieSpi2001,SanBenCal2006,GouSchHof2008,FLASH2007,NabHasTak2005, 
DroZepGop2006,NauNeeSok2004,SerYakSer2007,ZhaLytPop2007,NomHorTza2009,FenGilMas2009} offers the opportunity to investigate TPDI over a wide range of energies, from \emph{direct} (nonsequential) to sequential ionization. 
For photon energies below $I_2\approx54.4 \ev$, the second ionization potential of helium, one photon does not carry sufficient energy to ionize the residual $\He^+$ ion in its ground state.
After the first ionization event below this threshold, the second photoabsorption event has to occur within a short time interval such that the outgoing electrons can exchange energy for the double ionization process to take place. 

However, above this threshold, sequential double ionization (SDI) becomes possible: the first photon singly ionizes the helium atom and the second photon ionizes the remaining $\He^+$ ion, each of which constitutes a separate on-shell process.
The time interval elapsed between the two photoabsorption events and, thus, between the electron emission events, can be, in principle, arbitrarily long.
This sequential process can be qualitatively well described within an independent particle picture while quantitative details are influenced by electron-electron interactions \cite{PalResMcc2009,HorResMcc2008,GuaBarSch2008,HorMccRes2008,HorMorRes2007,FouLagEdaPir2006,BarWanBur2006,IvaKhe2009,LauBac2003,FeiNagPaz2009,FeiPazNag2009,IshMid2005}.

By contrast, electron correlation is a conditio sine qua non for the direct nonsequential double ionization (NSDI) process to occur. 
Therefore, much effort has been spent on investigations of electron correlation in the nonsequential regime.
So far, partially integrated quantities and total cross sections have been measured  \cite{NabHasTak2005,SorWelBob2007,AntFouPir2008,RudFouKur2008,KurFeiHor2010}.
Several theoretical studies of fully differential cross sections have been presented \cite{ColPin2002,HuColCol2005,HorResMcc2008,FeiNagPaz2008,GuaBarSch2008,PalResMcc2009}.
However, even on the level of total cross sections the different theoretical approaches lead to differences of more than one order of magnitude \cite{PalResMcc2009,FeiNagPaz2008,GuaBarSch2008,HorMccRes2008,HorMorRes2007,NikLam2007,FouLagEdaPir2006,NepBirFor2010}. 
The reasons for these discrepancies are still under debate, and the experiments could, up to now, not reach the needed accuracy to resolve them.
The extraction of cross sections close to the threshold for sequential ionization has remained a challenging problem.\\

The focus of the present communication is on the close similarity and underlying common features of the NSDI below the threshold $I_2$ and the SDI above the threshold $I_2$.
While the continuity across thresholds is appreciated as a consequence of the analyticity of the \emph{S-matrix} and is frequently involved in the determination of threshold exponents (Wigner and Wannier exponents \cite{Wig1948,Wan1953}), its implication for energy- and angular distributions in the TPDI process has, so far, not been explored.

We show that energy distributions and, to a considerable extent, also angular distributions of TPDI display universal features present both above and below the threshold for SDI.
These universal features become obvious when observables are analyzed in terms of the energy spacing (or asymmetry of energy sharing) $\DE=E_1-E_2$ of the two outgoing electrons.
The significance of this parameter controlling the double ionization irrespective of the value of $\hw$ can be understood within the framework of second-order time-dependent perturbation theory.
We will discuss the scaling of the resulting \emph{shape function} with pulse duration $T$.

Application to angular distributions characterized by anisotropy parameters $\beta_j$ shows that $\DE$ still plays a significant role even though the $\beta_j$ refer to reduced one-electron variables.
The latter are of experimental relevance, as a coincidence measurement is not required, at least for electron energies where the spectrum is not dominated by single ionization. We compare the single-electron angular distribution of TPDI with some results from earlier calculations \cite{BarWanBur2006,KheIvaBra2007b,IvaKhe2009}, and find significant differences.

Unless otherwise stated, atomic units will be used throughout the text.

\section{Method}\label{sec:method}

In our computational approach (see \cite{FeiNagPaz2008,Fei2009,SchFeiNag2011} for a more detailed description) we solve the time-dependent \Schro equation in its full dimensionality, including all interparticle interactions. We employ a time-dependent close-coupling scheme \cite{ColPin2002,LauBac2003,HuColCol2005,PinRobLoc2007} where the angular part of the wave function is expanded in coupled spherical harmonics. 
In order to reach convergence in the angular coordinates, we use single-electron angular momenta up to values of $\lonemax=\ltwomax=10$.
The highest total angular momentum included in the time propagation is typically $\Lmax=2$, which is sufficient since only the two-photon
channels $L=0$ and $L=2$ play a significant role. We checked explicitly that using $\Lmax=3$ does not change the results presented providing clear evidence that lowest-order perturbation theory in the photon field provides the dominant contribution.
For the discretization of the two radial variables, we employ a finite element discrete variable representation (FEDVR)~\cite{ResMcc2000,MccHorRes2001,SchCol2005,SchColHu2006}. 
The sparse structure of the resulting matrices allows for efficient parallelization, which is crucial to obtain results in the long-pulse limit (up to more than 20 femtoseconds propagation time). 
Radial boxes with an extension of up to $r_\submax=2000\au$ containing FEDVR elements with lengths of $4-4.4\au$ and of order $11$ are used for the presented results.
For the temporal propagation of the wave function, we employ the short iterative Lanczos method \cite{ParkLight86,SmyParTay1998,Lefo90} with adaptive time-step control. 

The laser field is assumed to be linearly polarized and treated in dipole approximation. The interaction operator is 
implemented in both length and velocity gauge, such that gauge independence can be explicitly verified.
We choose the temporal shape of the vector potential to be of the form
\begin{equation}\label{eq:lasfield}
\cvec{A}(t)=\unitv{z}A_0\sin^2\left(\frac{\pi t}{2\Tp}\right)\sin(\omega t)
\end{equation}
for $0<t<2\Tp$. 
The FWHM of the $\sin^2$ envelope function has the duration $\Tp$.
A peak intensity of $I_0=10^{12}\Wcm$ ensures that ground state depletion and three- or higher-order photon effects are negligible. 
 
The asymptotic momentum distribution is obtained by projecting the wave
packet onto products of Coulomb continuum states. 
These independent-particle Coulomb wave functions are not solutions 
of the full Hamiltonian. However, as we have previously demonstrated, projection errors can be reduced and controlled to the one-percent level by delaying the time of projection until the two electrons are sufficiently far apart from each other \cite{FeiNagPaz2008}.
All results were tested for numerical convergence and gauge independence.

\section{Shape function for two-photon double ionization}\label{sec:shapefunction}

The point of departure of our analysis of common features of TPDI above and below the sequential threshold observed in the numerical calculations is the spectral \emph{shape function} within second-order time dependent perturbation theory.
The applicability of the latter is indicated by the negligibly small contributions of three-photon (or higher-order) processes.
Similar approaches have been employed previously, see \cite{HorMorRes2007,PalResMcc2009} 
and references therein. 
To second order the amplitude of the transition driven by the xuv pulse is
\begin{multline}\label{eq:td_pert_2nd_transamp}
  t_{i\to f}^{(2)} = -\sumint{n} \Int_{t_0}^{t_f}\!\dt_1 \Int_{t_0}^{t_1}\!\dt_2 
  	e^{i E_{fn} t_1} e^{i E_{ni} t_2} \times\\
  	\times \bra{f} \op{V}(t_1) \ket{n}\bra{n} \op{V}(t_2) \ket{i} 
\end{multline}
with $E_{fn}=E_f-E_n$ and $E_{ni}=E_n-E_i$.
The transition probability for TPDI is 
$\PDI(E_1,\Omega_1,E_2,\Omega_2) = |t_{i\to f}^{(2)}|^2$,
where $\ket{i}$ is the ground state and $\ket{f} = \ket{E_1\Omega_1,E_2\Omega_2}$.
Insertion of the interaction operator in velocity gauge, 
$\op{V}(t)=(\pz{1} + \pz{2})A(t) \equiv \op\mu A(t)$, leads to a factorization of each term in the sum over intermediate
states into a spectral function $\G$ that just depends on the energies of the involved states and the temporal shape of the interaction potential, and a time independent matrix element depending on two dipole operators,
\begin{subequations}
\begin{gather}\label{eq:td_pert_2nd_transamp_2}
  t_{i\to f}^{(2)} = -\sumint{n} \braOket{f}{\op\mu}{n}\braOket{n}{\op\mu}{i} \G(E_{fn},E_{ni},\left[A(t)\right]) \\
\intertext{with}
  \G(E_{fn},E_{ni},\left[A(t)\right]) = \Int_{t_0}^{t_f}\!\dt_1 \Int_{t_0}^{t_1}\!\dt_2 e^{i E_{fn} t_1} e^{i E_{ni} t_2} A(t_1)\label{eq:shapeG} A(t_2) \eqstop
\end{gather}
\end{subequations}

Employing the rotating wave approximation and using the temporal shape of the pulse (\autoref{eq:lasfield}), \autoref{eq:shapeG} becomes
\begin{subequations}\label{eq:td_pert_g}
\begin{gather}
\begin{split}
  &\Gsinsq(\Delta\Efn,\Delta\Eni,A_0,T) = \\
&\quad\frac{A_0^2}{4} \Int_{0}^{2T}\!\dt_1 \Int_{0}^{t_1}\!\dt_2 F(\Delta\Efn,t_1,T) F(\Delta\Eni,t_2,T)\\
\end{split}
\intertext{with}
  F(\delta,t,T) = e^{i\,\delta\,t} \sin^2\left(\frac{\pi t}{2T}\right) \eqcomma
\end{gather}
\end{subequations}
where $\Delta\Efn=\Efn-\omega$ and $\Delta\Eni=\Eni-\omega$.
The latter variables denote the energy defects relative to the energy conserving \emph{on-shell} transition in each of the two steps.
The integral can be solved analytically, but is not shown here for brevity.

Above the sequential threshold ($\hw>I_2$) and for infinitely long pulses $\Tp\rightarrow\infty$ the angle-integrated joint transition probability $\PDI(E_1,E_2)$ is governed by $\G(\Delta\Efn=0,\Delta\Eni=0,A_0,T)$, \ie the on-shell part of the transition amplitude.
For short pulses and/or $\hw<I_2$, the behavior of $\PDI(E_1,E_2)$ is controlled by the deviations $\Delta\Efn\neq0$ and $\Delta\Eni\neq0$.
\autoref{eq:td_pert_2nd_transamp_2} can be further simplified by noting that among all possible intermediate states, the continuum states 
$\ket{Ep,1s}$ associated with the ionic ground state and an outgoing $p$-wave strongly dominate.
In other words, \emph{shake-up} or \emph{shake-off} processes only represent a few-percent correction in the SDI regime.

We further make the approximation that the transition amplitude from the intermediate to the final state 
is diagonal in the energy $E_1$ of the free electron, \ie that 
$\bra{E_1\Omega_1,E_2\Omega_2}\op\mu\ket{E_1'p,1s} \propto \delta(E_1'-E_1)$.
Consequently, a single term $\ket{n_0}\equiv\ket{E_1p,1s}$ in the sum over intermediate states provides the leading contribution.

Instead of using an independent-particle approximation for the transition matrix elements (as in \cite{HorMorRes2007,PalResMcc2009}), we approximate them by a constant.
This can be done if one is only interested in the shape of the final electron energy distribution. 
The transition probability to the final energies $(E_1,E_2)$ in this approximation
is given by
\begin{multline}\label{eq:tp_pert_2nd_transamp_flatapprox}
  \PDI_{\G}(E_1,E_2) = 
C|\Gsinsq(\Delta E_{fn_0^{(1)}},\Delta E_{n_0^{(1)}i},T) + \\
\Gsinsq(\Delta E_{fn_0^{(2)}},\Delta E_{n_0^{(2)}i},T)|^2 \eqcomma
\end{multline}
with $\Delta E_{fn_0^{(1)}}=E_f-(E_1+E^{(1s)}+\omega)$, $\Delta E_{fn_0^{(2)}}=E_f-(E_2+E^{(1s)}+\omega)$ and corresponding expressions for $\Delta E_{n_0^{(k)}i}$.

Accordingly, \autoref{eq:tp_pert_2nd_transamp_flatapprox} can be rewritten as
\begin{equation}\label{eq:tp_pert_2nd_transamp_flatapprox_de}
  \PDI_{\G}(E_1,E_2) = \PDI_{\G}(\DE,\Etot) \eqcomma
\end{equation}
with $\DE=E_1-E_2$. 
In the limit $T\rightarrow\infty$, the total energy of the final state $\Etot=E_1+E_2=2\hw+E_0$ is well determined by the photon energy and the ground state energy of helium, $E_0\approx-79\ev$.
For short pulses with finite Fourier width, the distribution of $\Etot$ is broadened accordingly.
The reduced probability density as a function of $\DE$ follows after integration over the Fourier width as 
\begin{equation}\label{eq:prop_dens_de}
\PDI_{\G}(\DE) = \frac{1}{2}\Int \PDI_{\G}(E_1,E_2)\dd\Etot
\end{equation}
(with a factor $1/2$ from the Jacobi determinant of the coordinate transformation).
For long pulses, this projection through the $(E_1,E_2)$ plane closely approximates the conventional one-electron energy distribution. \autoref{eq:prop_dens_de} has a distinct advantage when comparing pulses with different photon energies: the sequential peaks at $\hw-I_1$ and $\hw-I_2$ always show up at the same positions $\pm(I_1-I_2)$, irrespective of the photon energy.

We refer to \autoref{eq:tp_pert_2nd_transamp_flatapprox} as the shape function for double ionization. It contains both the spectral information of the \emph{time-independent} fully correlated two-electron Hamiltonian and the temporal structure of the pulse. 
It thus depends on the energy exchange between the two electrons included via ionization potentials that depend on the presence of the second electron, \ie $I_1\not=I_2$.
However, no additional information on electron correlations, in particular angular correlations, is included.

\begin{figure*}[tbp]
  \centering
  \includegraphics[width=\linewidth]{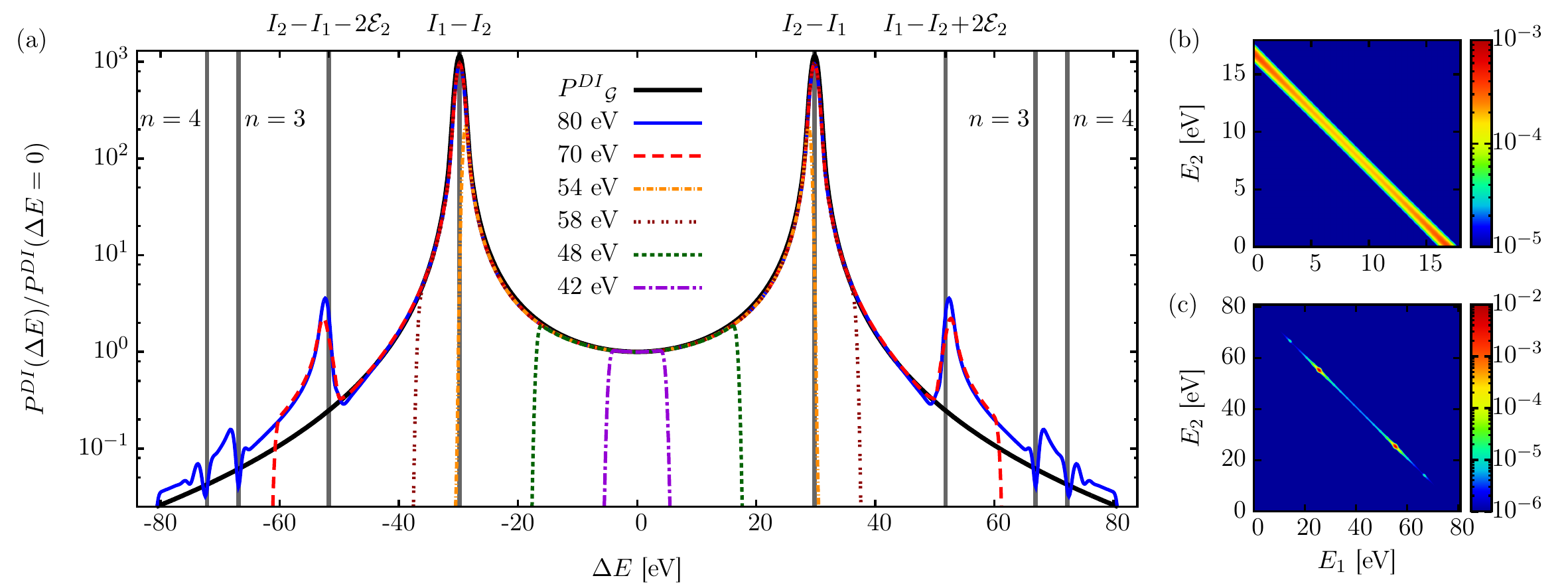}
  \caption{(a) Singly differential energy distribution $P^{DI}(\DE)$ as a function of the energy 
	difference $\DE=E_1-E_2$, normalized to the yield at $\DE=0$ for photon energies between $42\eV$ and $80\eV$.
	The pulses had a duration $\Tp=4.5\fs$ and an intensity $I_0=10^{12}\Wcm$. The gray lines show the expected positions of the peaks for the sequential process for different intermediate states (\ie with and without shake-up). Here, $\E_n$ is the excitation energy from the ionic $n=1$ state to an excited state $n$ ($\E_2\approx 40.8\ev$). The black line shows the energy distribution based on second-order perturbation theory $\PDI_{\G}(\DE)$. (b) and (c) show the two-electron energy distribution $\PDI(E_1,E_2)$ on the log scale for (b) $\hw=48\ev$ and (c) $\hw=80\ev$.}
  \label{fig:e1_hw_scan}
\end{figure*}

In \autoref{fig:e1_hw_scan} we compare $\PDI_{\G}(\DE)$ (\autoref{eq:prop_dens_de}) with the numerically exact solution of the time-dependent \Schro equation for a wide range of photon energies $42\ev\leq\hw\leq80\ev$ covering both the NSDI and SDI regimes with a pulse duration of $\Tp=4.5\fs$ for all pulses. 
For better comparison of the results at different photon energies, we have normalized each curve to the value at $\DE=0$. 

The excellent agreement with $\PDI_{\G}(\DE)$ for energy differences smaller than $\abs{\DE}\approx 45 \ev$ irrespective of $\hw$ exhibits the universal features present both in the SDI and NSDI process (note that the deviations visible for larger $\abs{\DE}$ result from the neglect of intermediate shake-up states in the sum in \autoref{eq:td_pert_2nd_transamp_2}).
Non-vanishing emission probabilities far away from the ``sequential'' peaks with $\DE_{fn}=\DE_{ni}=0$ signify energy sharing via electron-electron interaction irrespective of whether the sequential peaks are energetically accessible or not. 
Thus, also for pulses in the nominally sequential regime ($\hw>I_2$), a large part of the final-state space
can only be accessed through energy sharing that is governed by electron-electron interaction and thus closely resembles the NSDI process.
Only in the immediate vicinity of the on-shell peaks $\DE_{fn}=\DE_{ni}=0$, the width of which decreases as $T^{-1}$, SDI and NSDI processes become distinct.

\section{Angular distributions}\label{sec:angdis}

\begin{figure*}[t]
  \centering
  \includegraphics[width=\linewidth]{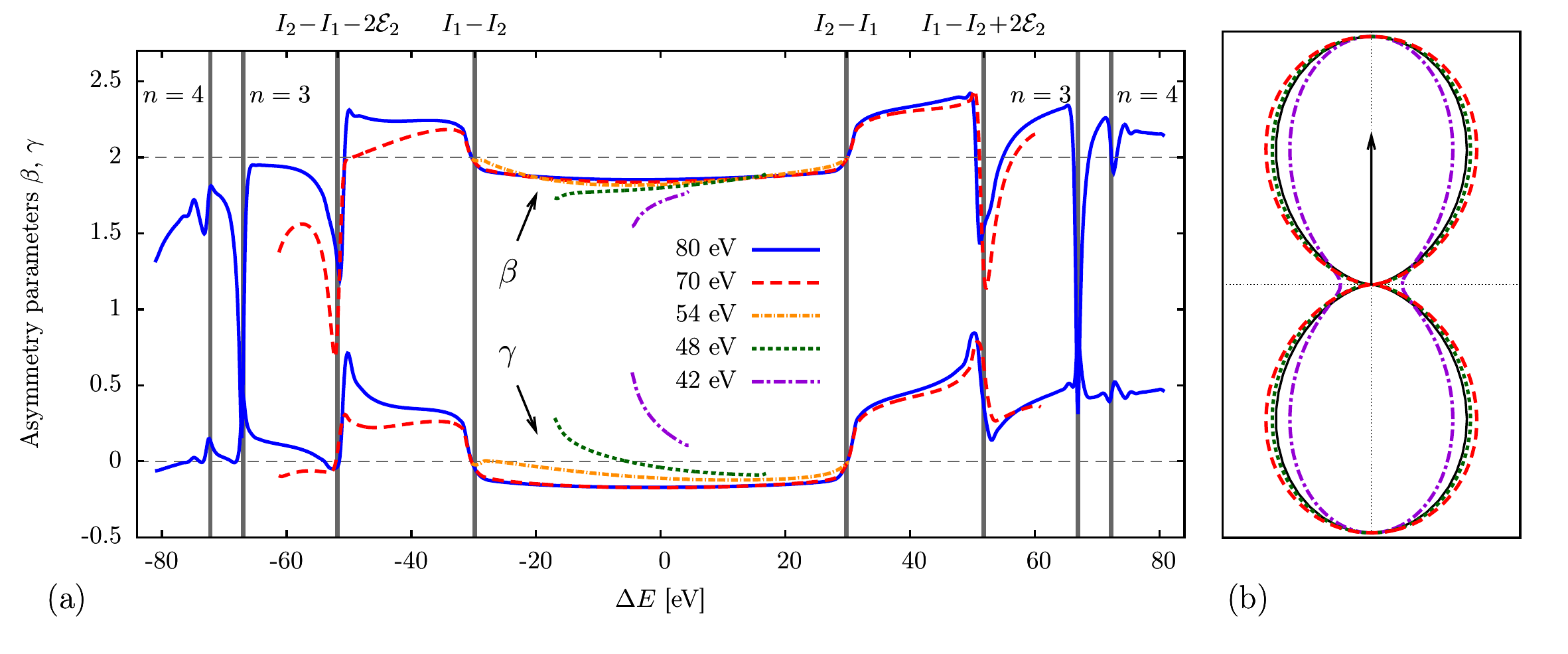}
  \caption{(a) Anisotropy parameters $\beta(\omega,\DE)$ and $\gamma(\omega,\DE)$ (see \autoref{eq:aspar2}) for photon energies between $42 \eV$ and $80 \eV$ as a function of the energy difference $\DE=E_1-E_2$ in the interval $\left[-E_{tot},E_{tot}\right]$, with $E_{tot}=2\hw+E_0$, $E_0\approx-79\ev$. The pulse parameters are the same as in \autoref{fig:e1_hw_scan}. The upper group of lines contains the $\beta$ parameters, while the lower group contains the values for $\gamma$. Note that the angular distribution of the electron with $E_1$ is shown, thus $\DE>0$ characterizes the faster electron and $\DE<0$ the slower one. (b) Angular distribution at equal energy sharing $\PDI(\DE=0,\theta)$ for $\hw=42\ev, 48\ev$ and $70\ev$ (from inside to outside, same color code as in (a)), normalized to a value of 1 for $\theta=0^\circ$. The laser polarization axis is indicated by the arrow. The thin black line shows a $\cos^2(\theta)$ distribution.}
  \label{fig:escan_bg}
\end{figure*}

We extend now the analysis of the DI probability as a function of $\DE$ to a second degree of freedom, the emission angle of one of the emitted electrons with respect to the laser polarization axis, $\theta$.
Conventionally the one-electron TPDI distribution $\PDI(E,\theta)$ is considered, which is obtained by integrating the full two-electron distribution $\PDI(E_1,E_2,\Omega_1,\Omega_2)$ over the energy and angle of one electron. 
This corresponds to measuring electrons without coincidence requirements.
Following our findings that the TPDI distribution depends primarily on $\DE$, 
we therefore investigate in the following $\PDI(\DE,\theta_1)$.
Two features are worth noting: since $\DE\approx2E_1-(E_0+2\hw)$, the switch from $E_1$ to $\DE$ corresponds to a coordinate shift in the limiting case of long pulses, \ie narrow Fourier width of $\hw$. 
Furthermore, since only the angle of one electron ($\theta_1$) is observed, $\PDI(\DE,\theta_1)$ is, unlike $\PDI(\DE)$, not symmetric under inversion $\DE\rightarrow -\DE$.

The one-electron distribution has cylindrical symmetry (\ie $\phi$ is cyclic), and can be parametrized in terms
of the anisotropy parameters $\beta_j$ obtained from expanding $\PDI(\omega,\DE,\theta_1)$ in Legendre polynomials
$P_l(\cos\theta_1)$,
\begin{equation}\label{eq:aspar}
\PDI(\omega,\DE,\theta_1)= \PDI(\omega,\DE) \sum_{j=0}^{\infty} \beta_j(\omega,\DE) P_j(\cos\theta_1) \ed .
\end{equation}
In the following, we will label the anisotropy parameters $\beta_j$,  $\beta=\beta_2$ and $\gamma=\beta_4$ (as \eg in \cite{BarWanBur2006}).
The integration over the angular part of one electron can be performed analytically.
For two-photon double ionization from the ground state, the coefficients of Legendre polynomials with $j>4$ vanish. 
In addition, odd multipoles $\beta_j$ vanish because of parity conservation. 
Consequently, 
\begin{multline}\label{eq:aspar2}
\PDI(\omega,\DE,\theta_1)= 
\PDI(\omega,\DE)\times\\ 
\times \lbrace 1 + \beta(\omega,\DE) P_2(\cos\theta_1) + \gamma(\omega,\DE) P_4(\cos\theta_1) \rbrace 
\end{multline}
where we have indicated the explicit dependence of the angular distribution and anisotropy parameters on the photon energy $\hw$.

For uncorrelated sequential emission with the ground ionic state as intermediate state, \ie if each electron independently absorbs one photon from a $1s$ state, we expect a dipole-like $\cos^2(\theta)$ distribution for both electrons.
This corresponds to $\beta=2$ and $\gamma=0$.
Deviations from this scenario, in particular correlated and temporarily confined joint emission manifests itself by deviations from these values.

In one-photon DI of He, for example, the photon can only be absorbed by one electron, while the second electron is mainly released due to shake-off.
Experiments (\cf \cite{BraDoeCoc1997} and references therein) showed that indeed $\beta$ is approximately zero for the slow electron, showing a clear sign of the isotropic shake-off process.
Note that $\gamma$ is always zero for one-photon processes. 
For $\hw$ close to the one-photon DI threshold ($\hw\geq79\ev$), $\beta$ is theoretically predicted to be close to $\beta\approx-1$ \cite{Gre1987} and also experimentally found to be negative (see \eg \cite{HueMaz2000} and references therein), indicating a preference for emission perpendicular to the polarization axis.

For two-photon DI a completely different scenario prevails.
For photon energies well below the threshold for sequential ionization, the energy distribution is flat, similar to that of the one-photon process. 
However, the angular distribution is completely different. This is to be expected because of the different number of photons absorbed, and, thus the different amount of angular momentum transferred.
The one-electron angular distribution at equal energy sharing, $\PDI(\DE=0,\theta_1)$, for different photon energies (\autoref{fig:escan_bg}b) closely resembles a Hertz-dipole distribution ($\sim\cos^2\theta$, thin black line) suggesting that each electron absorbs one photon. 
Although the electrons have to strongly exchange energy to reach equal energy sharing, this interaction does not leave a visible trace in the one-electron angular distribution.
The one-electron angular distribution does not provide detailed information on the correlated emission, unlike observables related to the joint two-electron distribution, such as the triply differential cross section (TDCS) (see \cite{FeiNagPaz2008,FeiNagPaz2009} and references therein), the forward-backward asymmetry between the electrons~\cite{FeiPazNag2009}, or the angle $\theta_{12}$  between the electrons (as will be discussed below, see \autoref{fig:escan_asym}).

The energy-differential one-electron angular distribution characterized by $\beta(\omega,\DE)$ and $\gamma(\omega,\DE)$ for different photon energies $\hw$ between $42 \eV$ and $80 \eV$ (\autoref{fig:escan_bg}a) displays a remarkably similar dependence on $\DE$ but, unlike the energy distribution, also noticeable variations with $\hw$.

Deviations from the uncorrelated limit ($\beta=2$, $\gamma=0$) are most pronounced for slow electrons ($\hw=42\ev$, with only $2.5\ev$ emission energy per electron above the double ionization threshold).
Hence, emission perpendicular to the polarization axis can be observed with $\gamma=0.6$ and $\beta=1.6$.
As mentioned above, the anisotropy parameters are \emph{not} symmetric relative to $\DE=0$.
The reason is that we observe the angular distribution of the ``first'' electron, $\PDI(E_1-E_2,\theta_1)$. 
For $\DE<0$, we observe the \emph{slower} electron, while for $\DE>0$ the \emph{faster} one.
The electron repulsion force, \ie the acceleration due to the interaction with the other electron, is the same for both electrons.
However, the relative momentum change due to the interaction is much larger for the electron with the smaller momentum.

For somewhat higher photon energies ($\hw\geq 48\ev$) the distribution mainly approaches the Hertz-dipole like $\cos^2\theta$ distribution with little difference between the NSDI and SDI regime highlighting the continuity across the SDI threshold also in the angular distribution. 
Furthermore, the distribution becomes approximately symmetric ($\DE\rightarrow -\DE$) for larger $\hw$.
In the sequential regime, an asymmetry in $\DE$ remains visible but confined to large asymmetric energy sharings ($\abs{\DE} \rightarrow E_{tot}$) where one electron is again slow. However, the ionization yields become very small for this case (see \autoref{fig:e1_hw_scan}).

We focus in the following on the anisotropy parameters for mean photon energies above the DI threshold ($\hw>79\ev$) and consider both ``long'' pulses ($T=4.5\fs$) and attosecond pulses ($T<500\as$).
For pulses of a few femtoseconds duration dynamically induced Fano resonances due to interferences between SDI via the ionic ground state and shake-up intermediate states are present \cite{FeiPazNag2009,PalResMcc2009}.
Note that in the limit of \emph{both} ultrashort and long pulses these interferences disappear as one of the two pathways becomes dominant 
(the nonsequential background for $T\to0$, the sequential shake-up peak for $T\to\infty$).

\begin{figure}[tb]
  \centering
	\includegraphics[width=\linewidth]{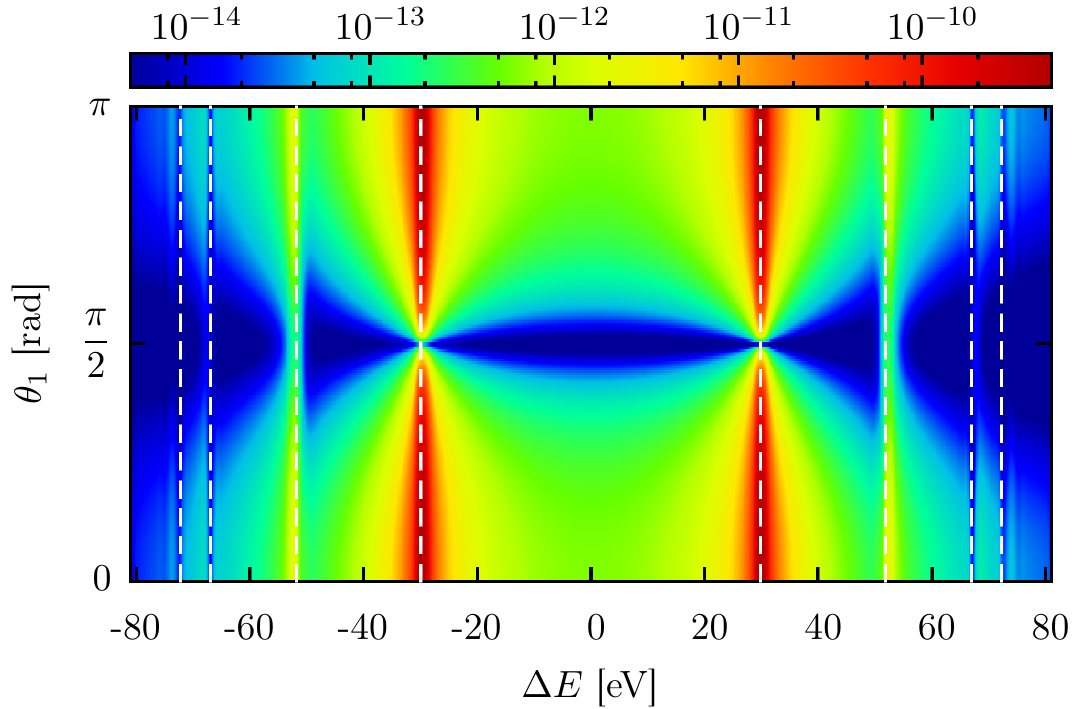}
   \caption{Two-dimensional energy-angular differential distribution  $\PDI(\DE,\theta_1)$ for an XUV pulse with $T=4.5 \fs$ and $\hw = 80 \ev$ as a function of the energy difference $\DE=E_1-E_2$ between the two electrons, and emission angle $\theta_1$, relative to the polarization axis. The vertical dashed white lines show the expected positions of the peaks for the sequential process, see \autoref{fig:e1_hw_scan}. The horizontal nodal line $\theta_1=\pi/2$ is visible except near shake-up resonances.}
  \label{fig:80ev_ang_escan}
\end{figure}

The Fano resonance structure is also visible in the one-electron angular distribution.
The two-dimensional one-electron energy- and angle-differential distribution $\PDI(\DE,\theta_1)$ (\autoref{fig:80ev_ang_escan}) reveals the resonant variation with $\DE$ at fixed $\theta_1$.
At $\theta_1=\pi/2$, the distribution has a narrow local maximum at energies where the sequential ionization process with shake-up in the intermediate state contributes.
Outside the shake-up resonance position, the persistence of the nodal lines of the Hertz dipole at $\theta_1=\pi/2$ (see \autoref{fig:escan_bg}b) is clearly visible.

Significant deviations can be only observed for energy sharings that correspond to resonant shake-up intermediate states ($n=2$, $n=3$) for which higher angular momentum ($l>0$) states can be populated. Thus, strong angular momentum mixing and deviations from an $s$-like initial (or in this case, intermediate) state can occur.
In turn, outside these shake-up resonances, the conservation of the nodal line of the Hertz dipole at $\theta_1=90^\circ$ is pronounced. 
This observation suggests that the $\beta$ and $\gamma$ parameters as a function of $\DE$ (or $E_1$, used in the following) are strongly interrelated.
Enforcing vanishing emission probability at $\theta=\pi/2$ implies
\begin{equation}\label{eq:asnod90}
\gamma'(E_1)=\frac{8}{3}\left( \frac{\beta(E_1)}{2}-1 \right) \ed .
\end{equation}

\begin{figure*}[tbp]
  \centering
	 \begin{minipage}{\linewidth}
	  \includegraphics[width=\linewidth]{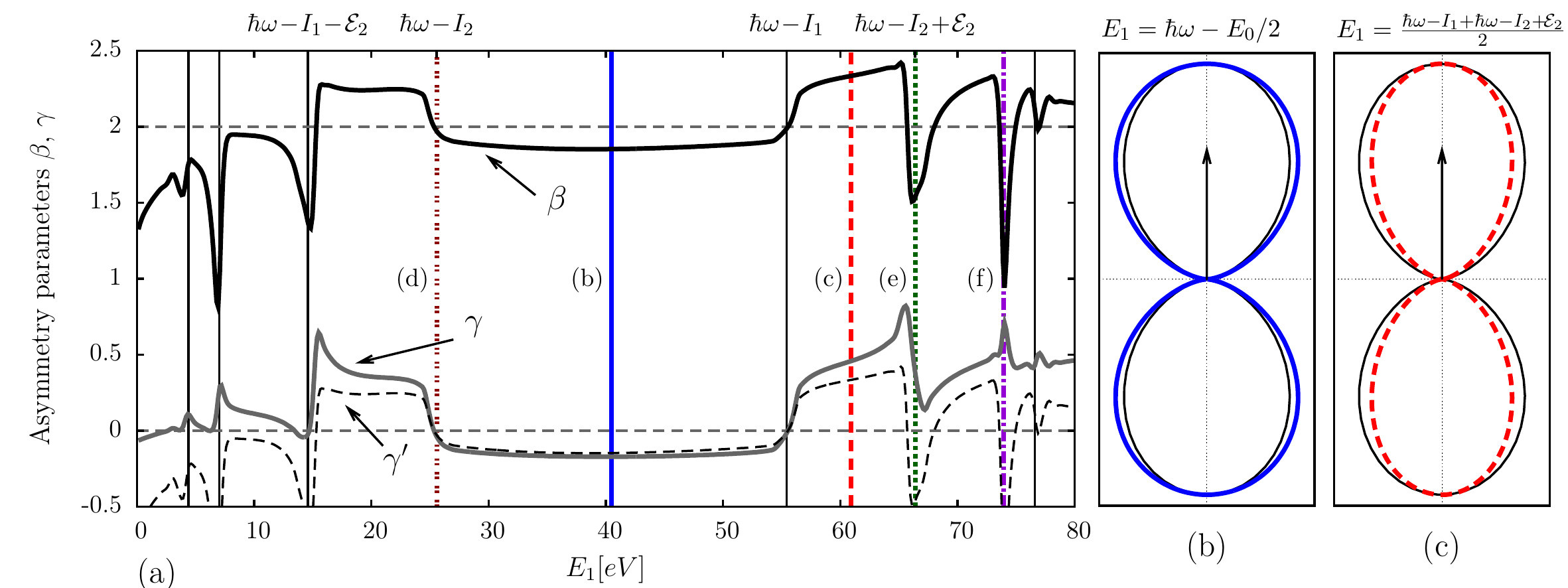}
	 \end{minipage}
     \begin{minipage}{0.45\linewidth}
	  \caption{(a) Anisotropy parameters $\beta(E_1)$ and $\gamma(E_1)$ as well as $\gamma'(E_1)$ (dashed line), which according to \autoref{eq:asnod90} enforces vanishing probability of electron ejection at $90^\circ$ to the polarization axis. (b)-(f) show angle-resolved one-electron distributions at different electron energies, normalized to a value of $1$ for $\theta=0^\circ$: (b) equal energy sharing, (c) asymmetric energy sharing, (d) main sequential peak (intermediate state: $\ket{\Hep1s}$), (e)  second sequential peak (intermediate state: $\ket{\Hep2l})$ (f) third sequential peak (intermediate state: $\ket{\Hep3l}$). Parameters: central frequency $\hw=80 \ev$, duration $\Tp=4.5 \fs$. The thin black line in (b)-(f) shows a $\cos^2$ distribution.}\label{fig:80eV_asym_param_corr}
     \end{minipage}
	\hspace{1cm}
     \begin{minipage}{0.45\linewidth}
	  \includegraphics[width=\linewidth]{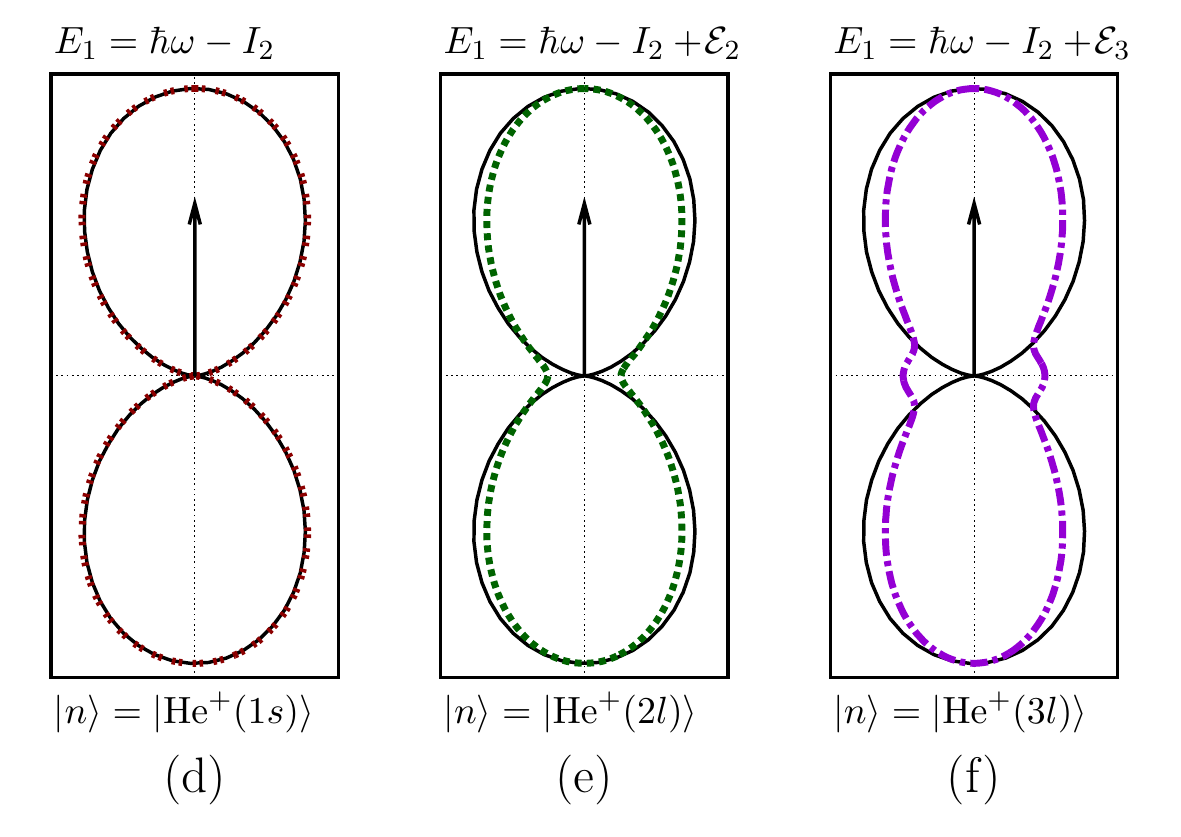}
     \end{minipage}

\end{figure*}

The symmetry-enforced value of $\gamma'(E_1)$ using the calculated values of $\beta(E_1)$ as input according to \autoref{eq:asnod90} agrees with the directly calculated $\gamma(E_1)$ remarkably well, in particular in between the $n=1$ sequential main peaks (\autoref{fig:80eV_asym_param_corr}a).
Exactly at the sequential peak we obtain $\beta=2$ and $\gamma=0$ as expected from the independent-particle process where the angular distribution is an uncorrelated product of $\cos^2\theta$.
Outside the main sequential peaks with increasingly asymmetric energy sharing the angular distribution becomes elongated along the polarization axis which is reflected in larger discrepancies between $\gamma'(E_1)$ and $\gamma(E_1)$.
The elongated emission pattern can be qualitatively explained by the fact that highly asymmetric energy sharing
can be reached through post-collision interaction (\cf\cite{BarBer1966,GerMorNie1972}), which is most effective when both electrons are 
emitted collinearly along the laser polarization axis.
The high-energy electrons are thus expected to belong to the subset of electrons at small angles,
explaining the elongated angular distribution.

For extremely unequal energy sharing, we find a striking change of the anisotropy parameters at the positions where sequential ionization through an intermediate shake-up state is possible (\autoref{fig:80eV_asym_param_corr}e,f).
These peaks and dips can be observed up to the $n=4$ shake-up with the current pulse parameters.
They are due to the fact that the intermediate ionic state for the sequential process can have higher angular momentum ($l\neq0$).
Hence for, \eg shake-up to $np$ states, the outgoing two-electron wave is not even approximately described by a $(p,p)$ wave, but is dominated by the $(s,d)$ contribution.
The interference between different intermediate state channels leading to the same final state gives rise to the complex angular distribution (\autoref{fig:80eV_asym_param_corr} (e) and (f) for the shake-up to intermediate states $n=2$ and $n=3$, respectively).\\

\begin{figure*}[tbp]
  \centering
  \includegraphics[width=\linewidth]{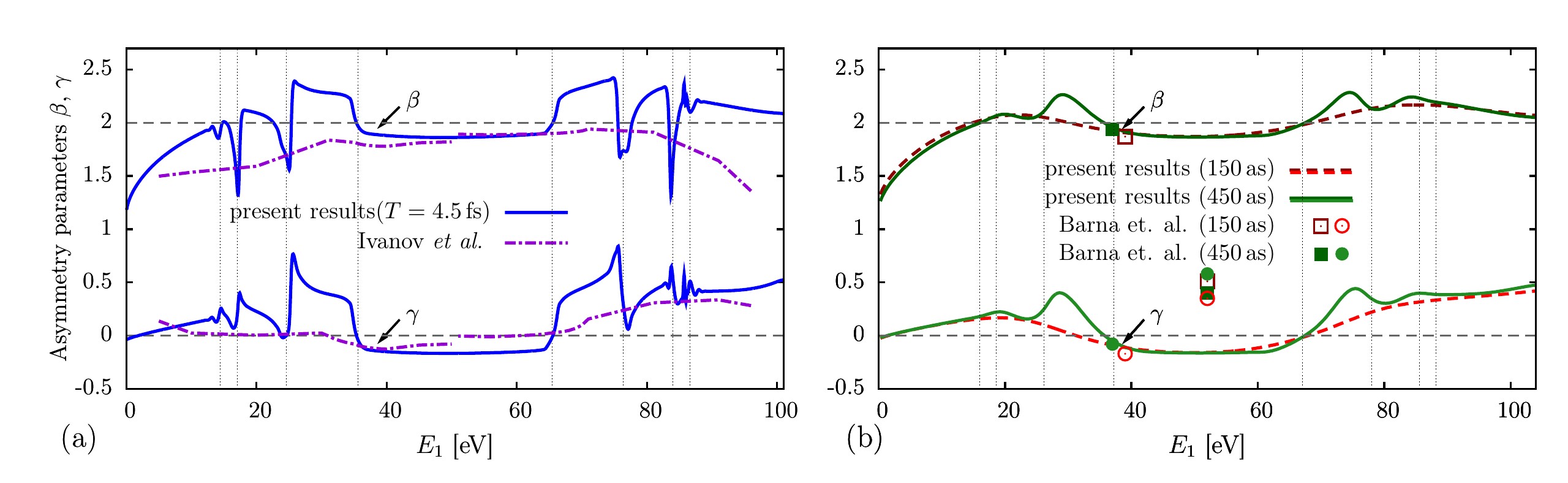}
  \caption{Asymmetry parameters $\beta$ (upper group of lines and squares) and $\gamma$ (lower group of lines and circles)
  for TPDI in the sequential regime with an XUV pulse with (a) $\hw=90 \ev$ in in the long pulse limit, $\Tp=4.5\fs$ (FWHM - $\sin^2$ pulse envelope) in comparison with recent time-independent results by Ivanov \etal \cite{IvaKhe2009}
  and (b) $\hw=91.6 \ev$ for ultrashort pulse durations, $\Tp=150\as$ and $\Tp=450\as$ (FWHM - Gaussian pulse envelope) in comparison with the calculations of Barna \etal \cite{BarWanBur2006}. }
  \label{fig:beta_time_comp}
\end{figure*}

The present calculation of anisotropy parameters for TPDI can be compared with two previously published results for long pulses \cite{IvaKhe2009} and for attosecond pulses \cite{BarWanBur2006}.
Comparison between the long-pulse limit of our present calculations (converged results are reached for $T=4.5\fs$) and recent time-independent results by Ivanov \etal \cite{IvaKhe2009}, (\autoref{fig:beta_time_comp}a), show reasonably good agreement near equal energy sharing.
However, because of the limitations and approximations in the convergent close-coupling (CCC) method \cite{IvaKhe2009}, such as the inequivalent treatment of the slow and fast electron, the intricate structures due to shake-up interferences appear to be missing.

For ultrashort attosecond pulses dynamical resonance structures are washed out. 
At the same time, the strong temporal confinement of the two electron emission events induces strong electron correlations and renders the distinction between sequential and nonsequential DI largely obsolete.
The measure of the temporal confinement to within a few hundred attoseconds on the angular distribution is therefore of particular interest.
We compare the one-electron angular anisotropy parameters for an XUV pulse with $\hw=91.6 \ev$ (previously investigated in \cite{LauBac2003,IshMid2005,BarWanBur2006}) and a pulse duration of $\Tp=150\as$ and $\Tp=450\as$ (FWHM of the Gaussian pulse envelope) with previous results by Barna \etal \cite{BarWanBur2006}. 
At the sequential peak we find rather good agreement (\autoref{fig:beta_time_comp}b). 
However, for equal energy sharing the calculation by Barna \etal overestimates correlation effects. This is probably due to two reasons: (i) the angular distribution in \cite{BarWanBur2006} was determined right at the end of the pulse when the electron momenta have not yet converged to their asymptotic values, and (ii) only single electron angular momenta up to $\lonemax=\ltwomax=2$ where included, which is not sufficient for converged angular distributions (see \cite{FeiNagPaz2008} for a detailed study of the convergence of TDCS with angular momentum).
The most striking difference is the absence of the nodal line at $\theta=\pi/2$ in \cite{BarWanBur2006} which we find to be well preserved for attosecond pulses.
The preservation of the nodal line suggests that for the limit of ultrashort pulses, TPDI proceeds by the absorption of one photon per electron as we have previously observed \cite{FeiNagPaz2009}.
Strong interaction between two almost simultaneously outgoing electrons leads to strong energy redistribution but leaves the nodal line in the angular distribution, to a good degree of approximation, intact.

\section{Angular correlation}\label{sec:fb_asym}

\begin{figure*}[tbp]
  \centering
  \includegraphics[width=0.95\linewidth]{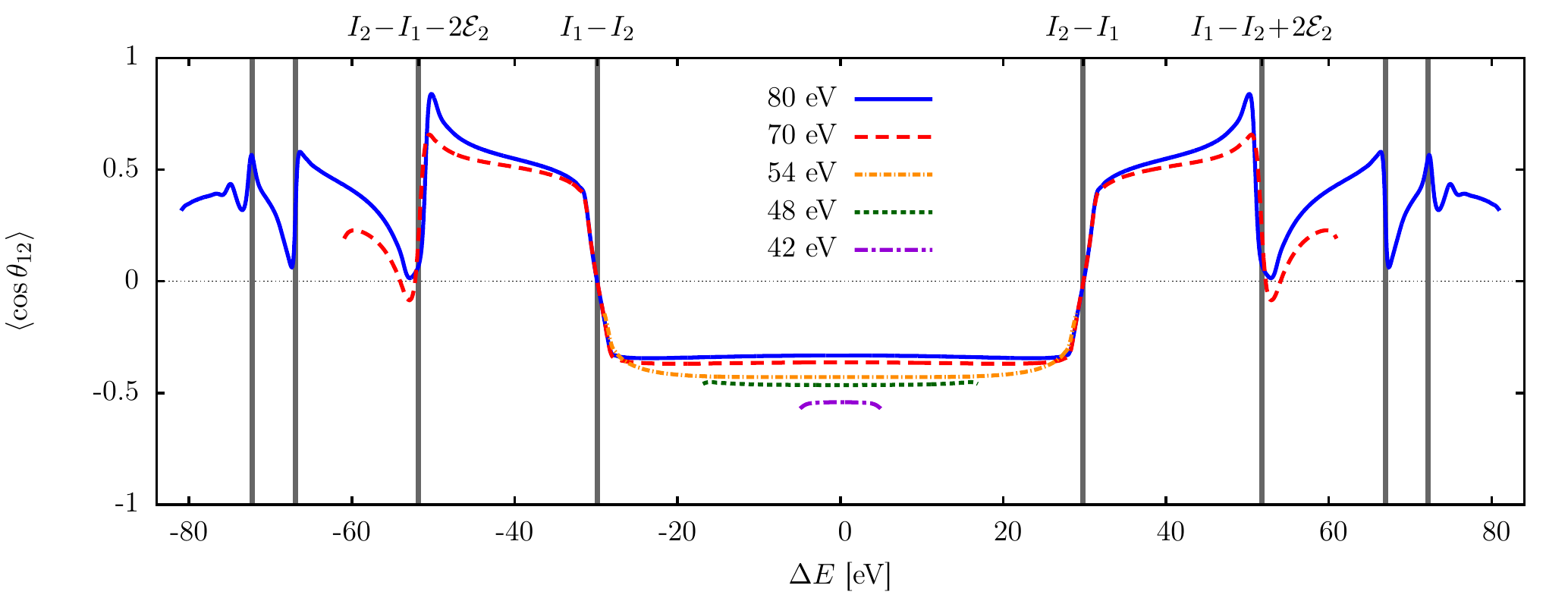}
  \caption{Angular correlation characterized by the expectation value of the angle between the emitted electrons, $\expcth(\DE)$, for photon energies between $42 \eV$ and $80 \eV$ (the other field parameters are the same as in \autoref{fig:e1_hw_scan} and \ref{fig:escan_bg}).}
  \label{fig:escan_asym}
\end{figure*}

Correlation effects in the spatial anisotropy of TPDI are expected to be more pronounced in two-electron observables derived from the joint probability density $\PDI(E_1,\Omega_1,E_2,\Omega_2)$.
Following the observation that the energy spacing $\DE$ governs the electronic dynamics both below and above the SDI threshold (\autoref{eq:tp_pert_2nd_transamp_flatapprox}, \autoref{eq:prop_dens_de}) we first integrate over $\Etot=E_1+E_2$ to arrive at the reduced probability $\PDI(\DE,\Omega_1,\Omega_2)$ as a function of the energy sharing $\DE$.
As a measure of the angular correlation between the two outgoing electrons we choose to investigate the angle $\theta_{12}$ between the electrons.
The expectation value $\expcth(\DE)$ provides a suitable quantity to measure the asymmetry in the joint angular distribution.
A negative value of $\expcth$ indicates back-to-back emission, while for positive $\expcth$ the electrons are emitted into the same hemisphere.

\begin{figure}[tbp]
  \centering
  \includegraphics[width=\linewidth]{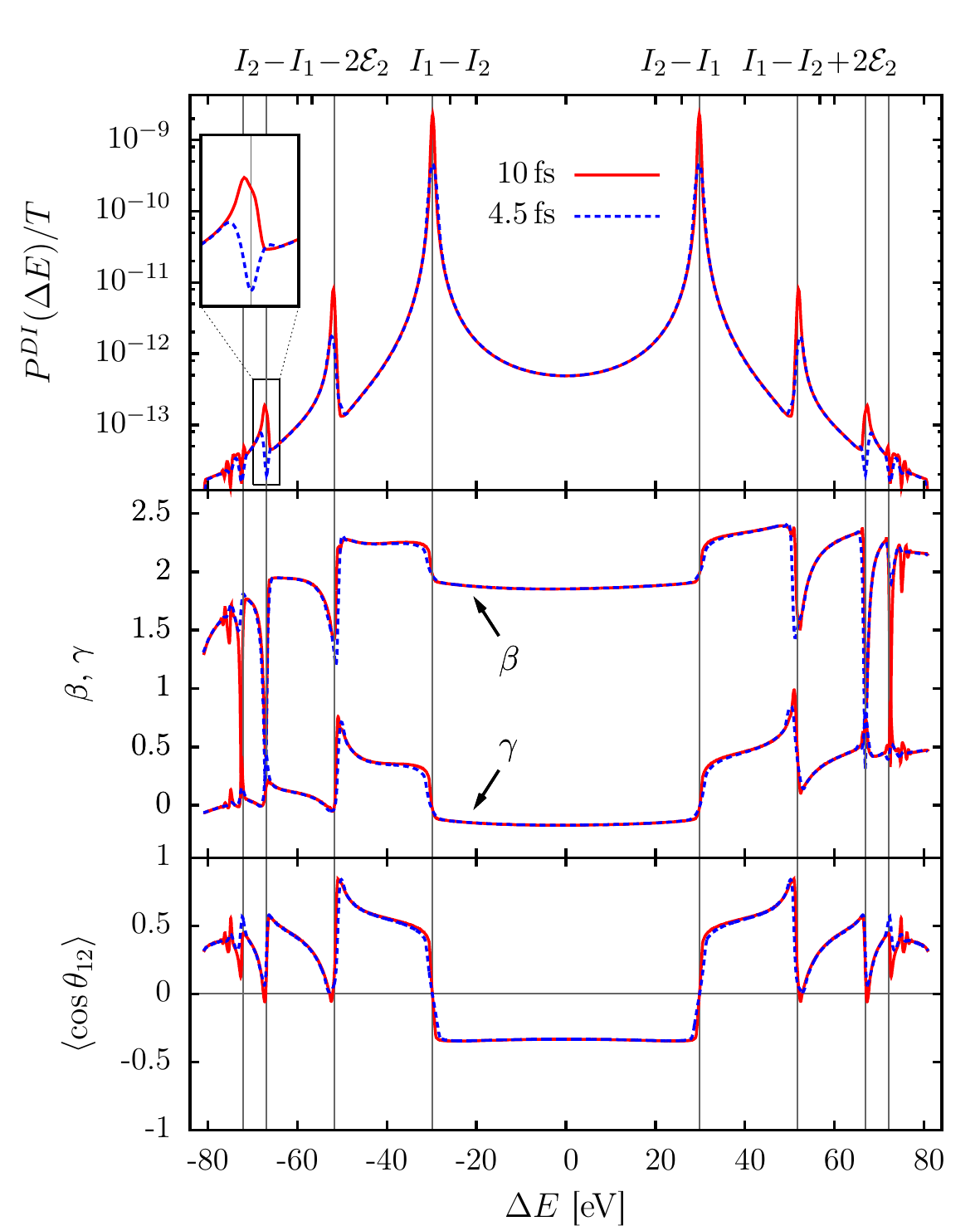}
  \caption{Energy distribution divided by the pulse duration $\PDI(\DE)/T$, anisotropy parameters $\beta$, $\gamma$, and $\expcth(\DE)$ as a function of $\DE$ for two different pulse durations: $T=4.5\fs$ (blue dashed line) and $T=10\fs$ (red solid line) and $\hw=80\ev$.}
  \label{fig:de_dist_10fs}
\end{figure}

This measure displays, as a function of $\DE$, very similar features over a wide range of photon energies, $42\ev-80\ev$, for a pulse duration of $T=4.5\fs$ (\autoref{fig:escan_asym}).
Regardless of the photon energy, the electrons are strictly emitted back-to-back between the main peaks of sequential ionization with the ionic intermediate ground state $\Hep(1s)$ and in the same direction for energy sharings outside the main peaks at $\pm(I_1-I_2)$.
For photon energies below the SDI threshold $\expcth(\DE)$ is strictly negative implying that electrons are predominantly emitted back-to-back in the NSDI regime.
Despite the apparent similarity of the $\DE$ dependence of the asymmetry (\autoref{fig:escan_asym}) and the anisotropy parameters (\autoref{fig:escan_bg}, \autoref{fig:80eV_asym_param_corr}) we emphasize one key difference: the behavior of $\expcth(\DE)$ is controlled by two-particle correlations while $\beta(\DE)$ and $\gamma(\DE)$ are one-electron variables.
Unlike the anisotropy parameters, in the strictly sequential regime $T\rightarrow\infty$ when the emission is confined to the peaks of on-shell ionic intermediate states, $\expcth(\DE)$ would approach zero.
The onset of this uncorrelated limit can be observed for moderately long pulses (\autoref{fig:escan_asym} and \autoref{fig:de_dist_10fs}):
at the energy difference $\DE$ corresponding to the sequential process with the intermediate ionic ground state the asymmetry goes to zero, $\expcth(\DE)=0$.\\

We conclude our discussion of the various distributions as a function of $\DE$ by comparing two different pulse durations in the long-pulse limit ($T=4.5\fs$ and $T=10\fs$) for a photon energy $\hw=80\ev$.
\autoref{fig:de_dist_10fs} shows the energy distribution divided by the pulse duration $\PDI(\DE)/T$, the anisotropy parameters $\beta$, $\gamma$, and $\expcth(\DE)$.
All three distributions show that convergence with pulse duration is already reached for $T=4.5\fs$ over a wide range of energy sharings. 
This underscores the linear scaling of the energy distribution for nonsequential ionization, \ie for all intermediate states that are \emph{off-shell}.
Only close to the peaks of the sequential process do the two distributions deviate from each other.
At the position of the resonances the yield scales quadratically with pulse duration, leading to a growth of the sequential contribution with respect to the nonsequential background.
An interesting consequence of this can be observed for the $n=3$ shake-up interference in \autoref{fig:de_dist_10fs} (see inset) where the dip in the Fano resonance for $T=4.5\fs$ turns into a peak for $T=10\fs$.

Similarly, the (joint) angular distributions are converged in the purely nonsequential regions. In contrast, the resonant features at the sequential peaks get more pronounced for the longer pulse since the sequential contributions scale as $T^2$ whereas the nonsequential background scales linearly with $T$.
For example, for a pulse duration of $T=10\fs$, $\expcth(\DE)$ approaches the long pulse limit $\expcth=0$ at the shake-up sequential peaks for $n=2$ and $n=3$, while it is still non-zero for the
shorter pulse.

\section{Total cross sections}\label{sec:angintcs}

We finally address the consequences of the continuity across the SDI threshold for angle-integrated cross sections for pulses with femtoseconds duration.
Starting point is the shape function (\autoref{eq:td_pert_g}, \autoref{eq:tp_pert_2nd_transamp_flatapprox}).
The continuity across the SDI threshold is explicitly incorporated by the assumption of constant (or smoothly varying) matrix elements (\autoref{eq:tp_pert_2nd_transamp_flatapprox}).
Accordingly, at a fixed value of $T$ the shape function is a smoothly varying function of the final-state energy $\Etot$ as well as of the photon energy $\hw$.
This is the reason underlying the notion of the ``virtual sequential ionization'' appearing in the vicinity but below the SDI threshold \cite{HorMorRes2007,HorMccRes2008}.
The only remaining manifestation of the SDI threshold in \autoref{eq:td_pert_g}, \autoref{eq:tp_pert_2nd_transamp_flatapprox} is the vanishing argument $\DE_{ni}=0$, $\DE_{fn}=0$ in the complex exponential functions when $\hw$ reaches the ionization potential of the ionic ground state, \ie the intermediate ionic ground state becomes on-shell.
As $\hw$ approaches this value from below, $\PDI(\DE)$ smoothly increases as $\DE\to\pm(I_2-I_1)$ and the energy mismatches $\DE_{fn}\rightarrow 0$ and $\DE_{ni}\rightarrow 0$ decrease.
The only apparent discontinuity between the nonsequential and sequential regimes appears in the limit $T\rightarrow\infty$ as the contribution of the shape function around the on-shell transition scales as $\sim\!T^2$, while all off-shell intermediate contributions for $\hw$ both below and above the SDI threshold scale linearly with $T$ as $T\rightarrow\infty$.
The width of the region around $\DE_{fn}=\DE_{ni}=0$ exhibiting quadratic scaling is proportional to $1/T$.
In the nonsequential regime only parts of the shape function with $\DE_{fn},\DE_{ni}\not=0$ are accessible. The total yield $\PDI=\int\PDI(\DE)\dd\DE$ thus scales linearly with $T$ for long enough pulses (\cf\cite{FeiPazNag2009}). 

Conventionally, one defines a generalized two-photon double ionization cross section as
\begin{equation}\label{eq:CS_def}
 \sigma_{(2)} = \lim_{T\to\infty} \frac{\omega^2}{I_0^2 T_{\text{eff},2}} \PDI \eqcomma
\end{equation}
with $T_{\text{eff},2}=35T/128$ for $\sin^2$ pulses and a two-photon transition.
\autoref{eq:CS_def} assumes that $\PDI$ is approximately constant over the spectral bandwidth of the pulse. 
At the SDI threshold the asymptotic scaling of $\PDI$ switches from $\sim\!T$ to $\sim\!T^2$ and $\PDI$ thus depends strongly on the (mean) photon energy.
The pulse should then be long enough to possess a sufficiently small spectral bandwidth to uncover the correct asymptotic scaling.
By the same token, however, the variation of $\PDI$ as a function of $\hw$ close to the threshold becomes more pronounced as the pulse duration becomes longer
and the approach to the underlying divergence of $\PDI(\DE)$ for $\DE\to\pm(I_2-I_1)$ is increasingly uncovered.
Therefore, the hallmark of the SDI threshold is a pronounced rise of $\PDI$ and, in turn, of $\sigma_{(2)}$.

An alternative way of extracting cross sections from a time-dependent calculation has recently been used by Palacios \etal\cite{PalResMcc2009}.
While it removes the uncertainty in the total energy of the pulse, the energy
resolution in the intermediate state is still determined by the pulse duration.
Close to threshold long pulses are thus still needed for accurate extraction of the cross section.
In addition, it requires very precise energy resolution in $\PDI(E_1,E_2)$.
For simplicity, we use \autoref{eq:CS_def} in the following.

\begin{figure}[tbp]
  \centering
  \includegraphics[width=\linewidth]{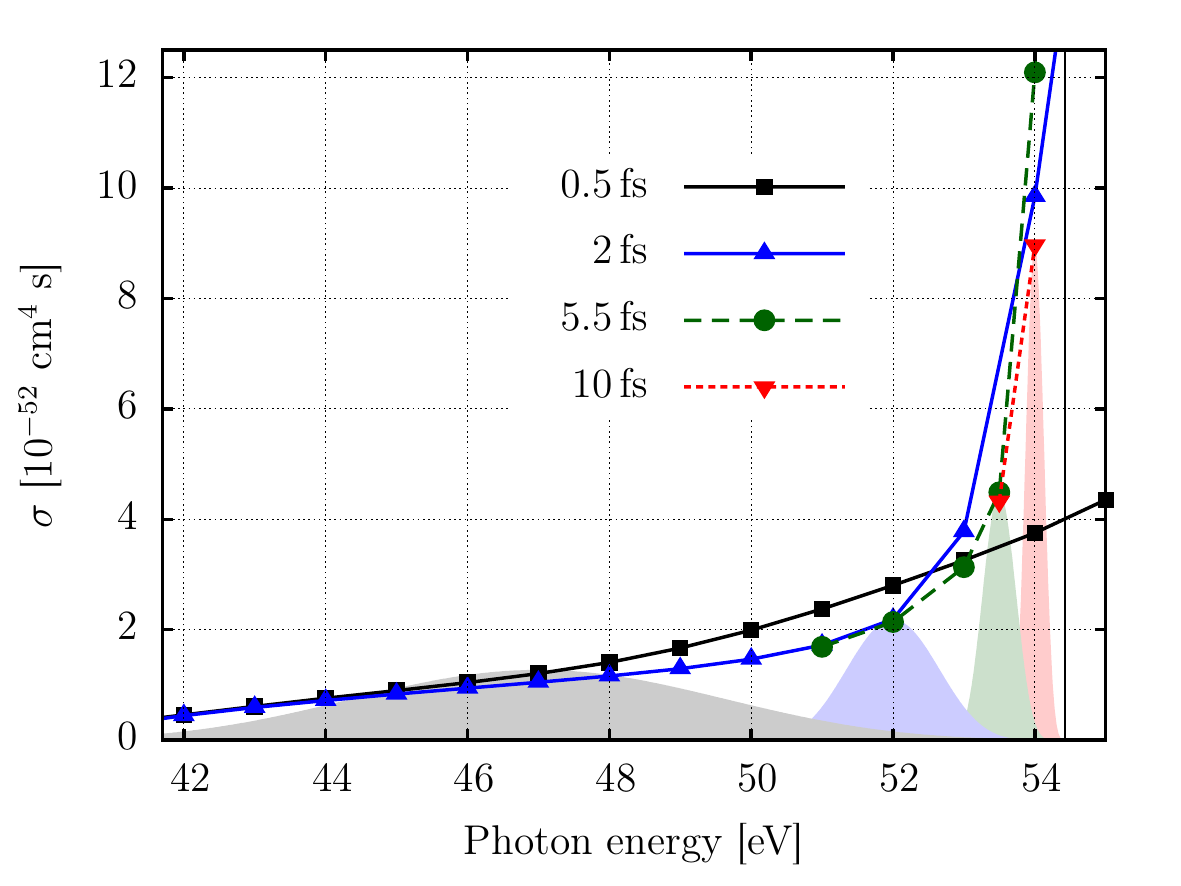}
  \caption{Total TPDI cross section obtained using $\sin^2$ pulses with different durations $\Tp$. Spectral distributions of the pulse are displayed for one data point for each $T$, showing the area over which the cross section obtained is effectively integrated. The intensity is $10^{12}\Wcm$ for all pulses. Angular momenta up to  $\Lmax=3$ for the total angular momentum and $\lonemax=\ltwomax=7$ for the single angular momenta are included. 
The radial boxes had extensions up to $r_\submax=1400\au$ for the $10\fs$ pulses.}
  \label{fig:nsTPDI_CS_thresh_beh}
\end{figure}

The total cross section for nonsequential TPDI strongly increases just below the SDI threshold, as anticipated above and observed in a number of recent studies \cite{HorMorRes2007, HorMccRes2008,FeiNagPaz2008,PalResMcc2009,NepBirFor2010}. 
This increase is a direct manifestation of the continuity across the threshold and can be qualitatively explained as follows: 
because of the time-energy uncertainty, the system can transiently ``borrow'' the energy that
is missing for the sequential pathway to be accessible, if the intermediate state (\ie the state after one-photon absorption) 
is only transiently occupied. 
In other words, while the final DI state is on-shell, the intermediate ionic state is off-shell but only barely so.
The occupation time of the intermediate state scales with the inverse of the borrowed energy.
For photon energies close to the SDI threshold, the system has to borrow only very little energy. 
In turn, the possible occupation time of the intermediate state corresponding to the sequential process increases leading to a large total yield. 
This process has previously been called ``virtual sequential'' TPDI \cite{HorMorRes2007,HorMccRes2008}.
Eventually, when the SDI threshold is crossed, both transitions can be on-shell and the intermediate state for the sequential process can be occupied for an infinitely long time.

In \autoref{fig:nsTPDI_CS_thresh_beh} we show the total nonsequential TPDI cross section (\autoref{eq:CS_def}), as extracted from pulses with a $\sin^2$ envelope and durations $\Tp$ from $0.5\fs$ (comparable to the frequently used 10 cycle pulses in literature) up to $10\fs$ (corresponding to a total pulse duration of $20\fs$). 
At some of the data points, we plot the spectral distribution of the laser pulses that were used to extract the cross section values.
Obviously, the frequency dependence of $\sigma_{(2)}$ near the SDI threshold can only be mapped out when the spectral width does not overlap with the above-threshold region.
Even for low photon energies the cross section is ``blurred'' when too short pulses are used.
This leads to a wrong slope of the cross section, as can be seen by comparing the results for $\Tp=0.5\fs$ with longer pulses.
\autoref{fig:nsTPDI_CS_thresh_beh} also demonstrates unambiguously that the rise of the cross section close to the sequential threshold does \emph{not} result from an unintended inclusion of sequential contributions to the total double ionization yield.
The spectral bandwidth of the $10\fs$ pulse is small enough ($\approx 0.3\ev$ FWHM) that virtually no contributions from the on-shell sequential process are present even at a central photon energy of $54\ev$.
The spectral contribution $\abs{\mathcal{F}(\omega)}^2$ with $\hw>I_2$ is only $0.026\,\%$ of the entire spectral content of the $10\fs$ pulse at $54\ev$ ($2.39\,\%$ for the $5.5\fs$ pulse).
From the value of $\PDI$ just above $I_2$ and the residual weight $\abs{\mathcal{F}(\omega)}^2$, the contribution to $\sigma_{(2)}$ at $54\ev$ can be estimated to be less than 0.42\%.

We can also rule out any contribution from the three-photon sequential process, since we only take partial waves with $L=0$ and $L=2$ into account when calculating $\sigma_{(2)}$. 
To verify that excitation of highly excited ionic states is not wrongly interpreted as double ionization,
we also checked the population in these states (which are occupied through two-photon single ionization). 
For the longest pulse used here ($10\fs$) and at a central frequency of $\hw=53.5\ev$, ions in the $n=1$ state after single ionization
can be resonantly excited into $n=7$ and $n=8$ by the second photon, whereas the population of higher $n$ states is strongly suppressed.
We have checked on the stability of both the ionic Rydberg population as well as the double continuum population as a function of propagation time after conclusion of the pulse and prior to projection for up to $3\fs$.
The temporal stability indicates that an unintended inclusion of high-lying Rydberg states in the calculated double ionization probability can be ruled out.

We therefore conclude that the rise of the generalized two-photon cross section as the SDI threshold is approached is a physical consequence of the continuity across the threshold and signature of the onset of the ``virtual'' SDI process.

\section{Summary}

We have discussed consequences of the continuity across the threshold from nonsequential double ionization (NSDI) to sequential double ionization (SDI) at $\hw=I_2$.
We have shown that the energy difference $\DE$ between the two outgoing electrons provides a suitable variable to display the common features of the NSDI and SDI processes on an energy- and angle differential level.

In particular, the singly differential energy distribution as a function of the energy difference between the electrons, $\PDI(\DE)$, agrees excellently over a wide range of energy sharings, irrespective of the photon energy.
The shape of the energy distribution can be understood by a simple model based on second-order time-dependent perturbation theory.

We have also presented, for the first time, fully converged anisotropy parameters for TPDI  over a wide range of photon energies ($42 - 80\ev$) and of pulse durations ($150 - 10000 \as$).
An approximately conserved nodal line at $90^\circ$ relative to the polarization axis, which remains undistorted by final-state electron-electron interaction, points to the fact that the ionization process is dominated by the absorption of one photon by each electron.
Deviations from this emission pattern could only be observed in regions where one of the electrons is so slow that its direction can be easily changed, \ie for very unequal energy sharing outside the main sequential peaks, or for low photon energies only slightly above the threshold for TPDI.
A second notable exception is the region of dynamical Fano resonances where additional intermediate pathways via excited ionic states open which can have non-zero angular momentum.
In such cases, the angular distribution does not even approximately resemble that of a Hertz-dipole shape.

Additionally, we have shown that a further consequence of the continuity across the threshold is a sharp rise of the energy- and angle integrated ``generalized'' cross section just below the threshold which can be viewed as the onset of the ``virtual'' sequential ionization channel consistent with the smooth approach to the on-shell intermediate state.

\begin{acknowledgments}
We thank I. Ivanov for sending us results in numerical form. R.P., J.F., S.N., E.P., and J.B.\ acknowledge support by the FWF-Austria, grants No.\ SFB016 and P21141-N16. JF acknowledges support by the NSF through a grant to ITAMP.
This research was supported in part by the National Science Foundation
through Tera\-Grid resources provided by NICS and TACC under grant TG-PHY090031.
Additional computational time was provided by the Vienna Scientific Cluster (VSC), and under Institutional Computing at Los Alamos National Laboratory.
The Los Alamos National Laboratory is operated by Los Alamos National Security, LLC 
for the National Nuclear Security Administration of the U.S.\ Department of Energy under Contract No.~DE-AC52-06NA25396.
\end{acknowledgments}

\end{document}